# The magnetic properties of a stack of elliptical nanomagnets for the initiation of apoptosis in cancer cells

Derek Michael Forrester

QinetiQ, Advanced Services & Products, Cody Technology Park, Farnborough, Hampshire, GU14 0LX

E-mail: **DMForrester@QinetiQ.com**

**Abstract**. The use of drugs or radiation to attack cancerous cells often leaves the treatment material within the body where it could affect healthy cells or damages healthy tissue; as is the case with traditional invasive surgery. An alternative approach is the creation of a nano machine that is directed to its target (the cancerous tissue), then activated to engage cancerous cells and 'surgically' induce controlled apoptosis. A magneto-mechanical device composed of nanomagnets that potentially meets this objective is thus predicated. The magnetic particles must be bio-functionalised for 'in-body' use where their uniqueness allows precise tracking and manipulation. In order to bio-functionalize magnetic particles they have to be embedded in a biocompatible coating or shell. Even enclosed within a biological entity the application of magnetic field gradients result in significant forces acting upon the magnetic particles. As such, single domain, ferromagnetic particles can be used inside the body for a number of applications. The lack of naturally occurring ferromagnetic particles inside the human body allows the bio-functionalized particles to be detectable, without the interference of noise from the host environment. Therefore, it is of paramount importance, for medical applications, to understand exactly how the magnetic particles will respond to a magnetic field gradient. Accordingly, the magnetic response of a system of four interacting magnetic particles is examined and it is shown that such a system can be optimised for targeting cancer cells and performing nano-surgical operations. The range of magnetic field required for such nano-medical applications is specified and the connection of these applications to different phase transitions arising in the magnetic field is described.

**1. Introduction**

Nanometre scale magnetic particles are increasingly being functionalized with biological applications in mind [1, 2]. Due to this, many studies are looking into increasing the tolerances of the human body to biofunctionalised nanoparticles [3]. In some of these studies it has been found that the geometry and size of the particles plays a significant role in determining whether a particle can overcome the physiological barriers it has to surmount inside the body [4]. When circulating in the body the particle has to evade the various uptake mechanisms that offer natural protection to the host. Evasion has been found to be optimised by designing particles that have elliptical rather than circular cross sections. Elliptical disk geometries have been shown to greatly increase the half-lives of particles as compared to spherical ones [4]. Internalization by phagocytosis of the nanoparticles was observed to occur quite rapidly when the elliptical disks approached a macrophage with their major axes perpendicular to the cell membranes [5]. However, when a disk approaches with its minor axis perpendicular to the cell membrane, the macrophage takes a much longer time to internalize the particle. Thus, any design process for biofunctionalisation of magnetic particles must consider the anisotropy of the nanoparticle with great care. The aspect ratio of the length to breadth of a particle dictates its successful navigation through the body [3].

Also, in order for a magnetic particle to survive inside the body it will usually have to be coated with biologically compatible substances. It is a subtle balance that has to be found between biological mimicry and functionality of the nanoparticles and this is reflected in the design methodology. The recent use of "magnetic viruses" is a prime example of this [6]. The magnetic virus is created by first extracting the DNA strands from inside a virus capsid (outer shell) leaving a biological stamp. Then multilayers are deposited into the empty ghost phage (virus shell). Thus, the magnetic virus is a hybrid of natural and man-made design that offers biocompatibility. This is but one example of reproducibly creating magnetic multilayers of uniform size and geometry. There are a number of other technologies available for producing monodisperse magnetic particles. Amongst these are the film-stretching technique [7] and nanotemplating using nanoimprint lithography [8]. The imprinted templates can be used for precise thin-film deposition of multilayer magnetic disks that can be used to create structures for specific medical applications.

In the human body natural cells are designed for specific purposes and possess unique shapes and mechanical properties. To work in harmony with the body a cell has to have the right size, shape, mechanical properties, surface texture and compartmentalization for its specific purpose [4]. With the design prerequisites for shape and geometry in mind, the magnetic properties of a stack or chain of magnetic particles are described in detail herein. Four magnetic particle interactions are examined. In comparison to the two [9-13] and three [14] magnetic particle multilayer stacks often studied, there is a quite different response to an applied magnetic field. The torque generated through the application of a magnetic field gradient is shown to be sufficient to initiate apoptosis in cancer cells.

It is proposed that a multilayer of single domain nanoparticles offers a high moment, low remanence alternative to superparamagnetic nanoparticles. These so called synthetic antiferromagnetic nanoparticles (SAF) have been shown experimentally to be amenable to functioning inside the human body [8]. Magnetic nanoparticles that possess high moments are useful for bio-magnetic sensing. The long range interaction of the magnetic particles with an external magnetic field allows the sensitive detection of the particles and also their manipulation. These magnetic nanoparticles should also possess low remanence. That is, after magnetically saturating a magnetic particle with an external magnetic field, the residual magnetisation should ideally be zero. Thus, the magnetic particles appear non-magnetic in the absence of an external magnetic field. This helps stop the aggregation of the nanoparticles due to magnetic attraction when particles are suspended in a colloid [15]. This is essential to prevent the nanoparticles from blocking blood vessels and causing embolisms.

Magnetic nanoparticles that exhibit superparamagnetism have many of the qualities desirable for medical applications [1]. One such quality is that superparamagnetic behavior occurs at or around room temperature. However, the main benefit of using superparamagnetic particles is that they have a very sensitive response to an applied magnetic field, which is much higher than conventional nanoparticles. Here we describe an alternative – high moment ferromagnetic multilayers separated by non-magnetic spacer layers. The size of these stable single domain ferromagnetic particles is larger than that of the very small superparamagnetic particles. As such, they have a larger net magnetic moment. This makes them ideal for a range of bio-medical applications. In particular, we specify a new device based upon the SAF structure that responds with enough magnetic torque to mechanically induce cellular death in cancer cells.

## 2. The magnetic properties of the stacked SAF

*2.1. The energy landscape*
The fabrication of monodisperse (size variations $<10\%$) ferromagnetic multilayer nanoparticles has recently been achieved [2]. These monodisperse CoFe layers are coupled antiferromagnetically through a Ru layer to produce zero remanence. The magnetic properties of two magnetic particles are described in our previous work and by others [9-13]. Here four magnetic particles are shown to also produce zero remanence and a hysteretic dependence upon the applied field strength and angle. To describe the magnetisation dynamics of the system we employ the Landau-Lifshitz-Gilbert equation,

$$\frac{d\mathbf{M}}{\gamma dt} = -\mathbf{M} \times \mathbf{H}_{eff} - \frac{\alpha}{M_S}\left[\mathbf{M} \times \left[\mathbf{M} \times \mathbf{H}_{eff}\right]\right] \quad (1)$$

where $\mathbf{M}$ is a magnetisation vector, $\gamma$ is the gyromagnetic ratio, $\alpha$ is the Gilbert damping factor, and $\mathbf{H}_{eff}$ is an effective magnetic field,

$$\mathbf{H}_{eff} = \frac{1}{\mu_0 V}\frac{\partial E}{\partial \mathbf{M}}$$

Where $E$ is an anisotropic Heisenberg Hamiltonian [cite PRB 2007]. We substitute into equation (1) the effective magnetic field of equation (2), $\mathbf{M} = M_S \mathbf{S}$ (where $M_S$ is saturation magnetisation), $E/\mu_0 V = \varepsilon M_S^2$, and $\tau = \gamma M_S t$, giving

$$\frac{d\mathbf{S}}{d\tau} = \mathbf{S} \times \frac{\partial \varepsilon}{\partial \mathbf{S}} - \alpha\left[\mathbf{S} \times \left[\mathbf{S} \times \frac{\partial \varepsilon}{\partial \mathbf{S}}\right]\right]$$

in dimensionless form. The system is described by the anisotropic Heisenberg Hamiltonian [16] which is reduced to

$$E = -J_{i,i+1}M_S^2 \sum_{i=1}^{3}\cos(\varphi_i - \varphi_{i+1}) + 0.5\mu_0 M_S^2 v\Delta D\sum_{i=1}^{4}\sin^2\varphi_i - g\mu_0 vH_a\sum_{i=1}^{4}\cos(\varphi_i - \beta) \quad (1)$$

The saturation magnetisation and volume are $M_S$ and $v$ respectively. The angle of the magnetisation vector in particle i with respect to the easy axis is $\varphi_i$. The energy associated with uniaxial shape anisotropy is $E_K = \mu_0 M_S^2 v\Delta D/2$, where $\Delta D$ is the difference between demagnetisation factors, $D_y - D_x$. The demagnetisation factors can be calculated using the methodology of Beleggia et al [17]. The last energy term in (1) is the Zeeman energy where $H_a$ is the applied magnetic field strength. The magnetic field is applied in the $x-y$ plane at angle $\beta$ to the easy axis. To simplify numerical calculations the energy equation (1) is now normalized by $|J|$ with $J = J_{i,i+1}M_S^2$. The coupling between each nearest neighbour is assumed to be the same. The notation is changed to accommodate this, with $H = \mu_0 vH_a M_S/|J|$, $K = \mu_0 M_S^2 v\Delta D/|J|$ and $\alpha = \pm 1$. This leads (1) to be rewritten in the form,

$$E = \alpha\sum_{i=1}^{3}\cos(\varphi_i - \varphi_{i+1}) + \frac{K}{2}\sum_{i=1}^{4}\sin^2\varphi_i - H\sum_{i=1}^{4}\cos(\varphi_i - \beta) \quad (2)$$

The sign of $\alpha$ is dependent upon whether the coupling is antiferromagnetic, $J_{i,i+1} < 0$ or ferromagnetic, $J_{i,i+1} > 0$ at $H = 0$. At a relatively large separation of the particles an antiferromagnetic coupling constant $J_{i,i+1}$ comes directly from the dipole-dipole interaction. These SAF multilayers consist of stable single domain nanoparticles and form the magnetic element of a biofunctionalised nanomagnetic system. The magnetic properties of the SAF will be discussed along with its ability to initiate programmed cell death in cancerous tissue. It will be shown that the theoretical model compares well with experimental data and that the SAF consisting of four magnetic discs can generate enough torque to mechanically begin the cancer cells destruction. The size of the SAF is between 30nm to 100nm in its longest dimension, making it highly suitable for use inside the body. Recently 30nm biofunctionalised nanoparticles have crossed the blood brain barrier [18].

*2.2. Phase diagrams and magnetic hysteresis*
In essence the interlayer between the magnetic particles in the nanomagnetic system acts as an artificial domain wall [16]. Separation of the magnetic particles by non-magnetic spacer layers of sufficient thicknesses enables the system of four magnetic particles to align antiferromagnetically with each other at $H = 0$. This is due to the magnetisations in the four artificial domains trying to minimize their magnetostatic energy interaction. The application of a magnetic field gradient to the system gives rise to a complicated evolution in the energy landscape. Stable minima and unstable maxima emerge in response. The stable and metastable states in these landscapes are found by solving five coupled non-linear equations: $dE/d\varphi_{1,2,3,4} = 0$ and the Hessian matrix set to zero. At all values of applied magnetic field strengths and orientations there exist stable energy minima in the energy landscape. In Figure 1 the critical lines of stability whereby the stable orientations of the magnetic moments at specific applied field strengths undergo a continuous or discontinuous phase change are demonstrated.

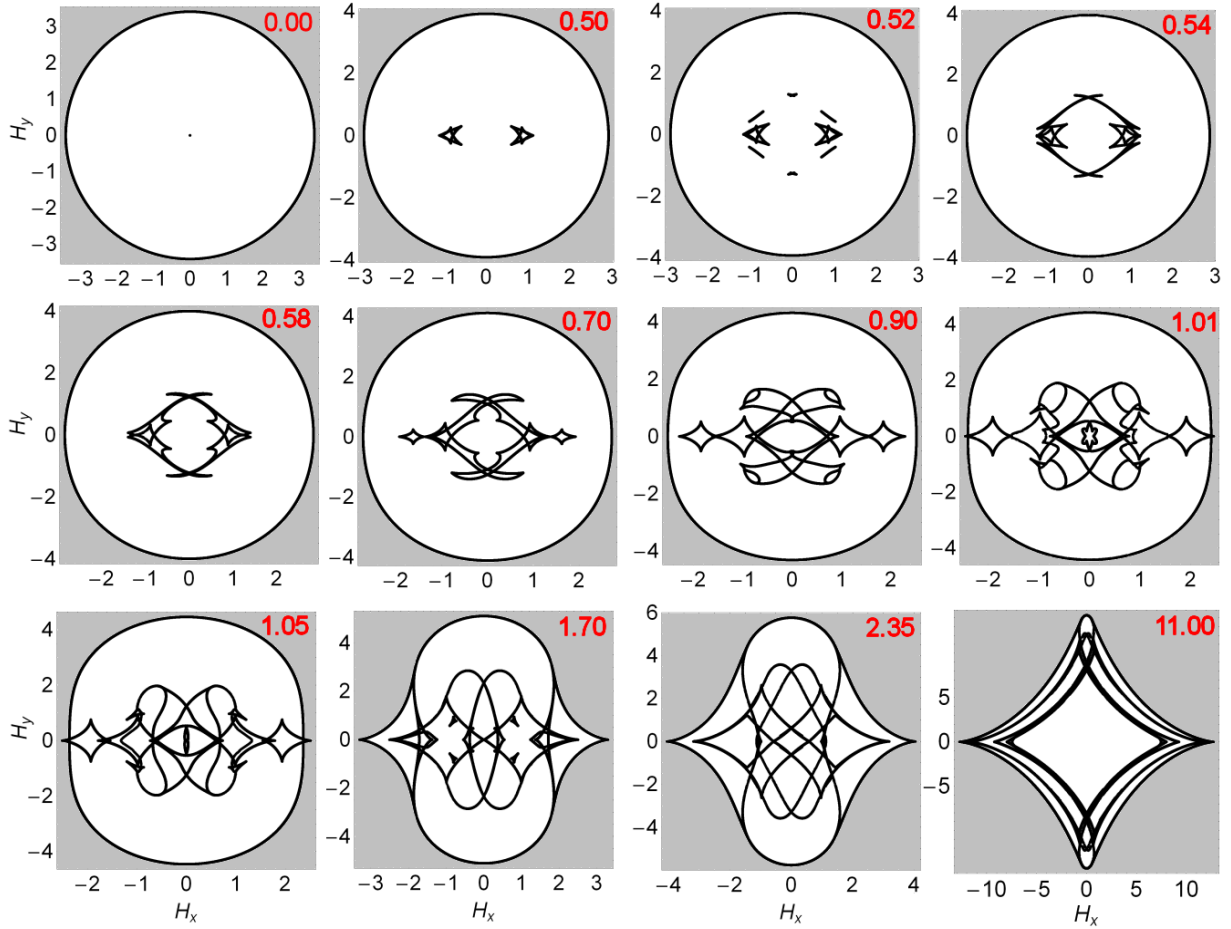

*Figure 1. The phase diagrams of four interacting magnetic elliptical disks such as in a synthetic antiferromagnetic multilayer structure, e.g.[CoFe/Ru/CoFe]3/Ru, are shown. The structure is subjected to an applied magnetic field in the $H_x - H_y$ plane of the magnetic disks. As the total energy of the system evolves in response to H, energy minima coalesce into saddle points, become deeper or shallower or emerge. The shape of the energy landscape is dictated by the interaction coupling between the magnetic particles, the strength and direction of the applied magnetic field, and the anisotropies of the disks. The angles, $\varphi_i$, associated with the magnetisation vectors of the magnetic disks change in response to the applied field and consequently the energy landscape shifts. Thus, the coupled four particles magnetisation angles rotate until the magnetic field is strong enough to give them all the same orientation. When this occurs they are ferromagnetically aligned. Ferromagnetic alignment is shown by the gray areas in the above plots. Before this occurs there is a complicated series of angle alignments as the magnetic field approaches saturation. All areas in the plots are related to stable orientations of the magnetisation vectors. The black lines mark the critical points where the angles of the magnetisation vectors change very rapidly to produce discontinuous phase transitions in the net magnetisation. They can also mark transitions to a new phase that occurs smoothly; a continuous phase transition. The plots show the critical lines of stable phase transitions for all angles of applied magnetic field.*

The magnetisation as a function of the applied magnetic field strength is shown in figure 2. The total magnetisation, M, is an average over the magnetic moment directions of the individual magnetic particles with respect to the magnetic field orientation:

$$M = \frac{M_s}{4} \sum_{i=1}^{4} \cos(\varphi_i - \beta) \qquad (3)$$

The system of magnetic and non-magnetic layers can be thought of as a unit of $N \leq 4$ domains. When the magnetic moment directions of each individual particle are the same, there is one domain. When each magnetic element has differing magnetisation vectors there are four domains. Figure 2 illustrates the sensitivity of the domain structure to the control parameters $H$, $K$ and $\beta$. Variations in these control parameters change the energy balance so that the domain structure is altered. In figure 2, the magnetisation is investigated in a system of particles with $K = 0.70$. For a synthetic antiferromagnetic multilayer it is clear that hysteresis occurs only at certain angles of the applied magnetic field. In figure 2 (a), the magnetic field is applied along the easy-axis direction of the stack of particles. The system shows hysteresis in its magnetisation as it responds to the applied field.

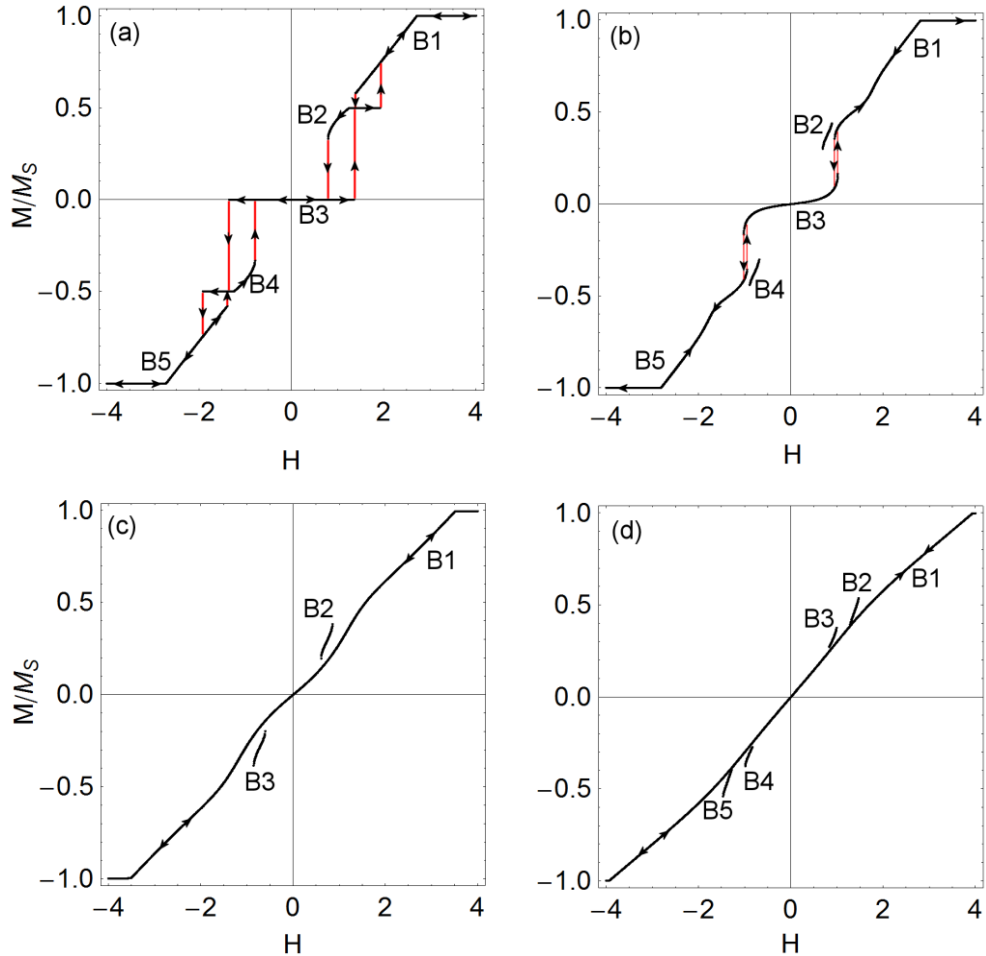

*Figure 2. For a system of four interacting magnetic particles there is a dependence of the magnetisation on the magnetic field H and the angle at which it is applied. The particles here have anisotropy parameter $K = 0.70$. (a) The magnetic field is applied along the longest axis of the multilayer stack, i.e. $\beta = 0°$. Four hysteresis loops emerge when the value of the magnetic field strength is cycled as $H = 0 \to 4 \to -4 \to 0$. The magnetic moment orientations of the four particles start at $H = 0$ with antiferromagnetic alignment. (b) At $\beta = 18°$ there exists two small hysteresis loops when starting with the same initial conditions as in (a) and the same cycle of H. (c) There is no*

*hysteresis to be seen when following the same magnetic field cycle as in (a) and (b) for $\beta = 54°$. There are, however, two new branches B2 and B3 that may require a fast cooling technique, as outlined in [16, 19], to be found experimentally. On these branches the magnetic moments are all directed in different orientations. They exist as a result of the complicated energy landscape. (c) Starting from antiferromagnetic alignment and $H = 0 \rightarrow 4 \rightarrow -4 \rightarrow 0$ there is no hysteresis for $\beta = 72°$. However, there are four stable branches B2-B4 that occur. They are punctuated at their end points by discontinuities which mark the points where the energy minima coalesce with saddle points. The real value of the magnetic field is found by multiplication of H by $|J|/\mu_0 \nu M_S$.*

The magnetisation in figure 2 shows the character of the system as it responds to the magnetic field. For the applied magnetic field angles that give rise to hysteresis, a cycled magnetic field destabilizes the energy landscape. This results in a series of Barkhausen jumps. When the magnetic field is slowly driven up with a constant sweep rate a slight change in the energy landscape occurs. Some energy minima become deeper and others become shallower. These decreasing energy wells eventually disappear at some critical field. Whenever the branch of magnetisation loses stability the system will instantly relax or jump to a still existing local minimum. For a broad range of applied magnetic field angles the magnetisation evolves continuously, i.e. without any discontinuous jumps, as the magnetic field changes. This is seen in figure 2 (c) and (d). However, there are some branches in these diagrams where discontinuities are seen. For example, in figure 2 (d) there are 4 extra branches B2-B5 which occur at specific levels of magnetic field strength. The interaction of the magnetic particles and their response to the magnetic field is intrinsically history dependent. Small changes in field sweep rate parameters can completely alter the pattern of discontinuities. As such it should be possible to find two dissimilar pathways through the energy landscape for similar magnetic field histories. Small changes in the sweep rate can revise the entire Barkhausen jump pattern. When the magnetisation approaches saturation the history of the system configuration is lost. Therefore reproducibility of the Barkhausen pattern should be difficult for large systems of particles. For smaller systems, it should be possible to replicate these patterns. Cooling the system in a magnetic field corresponding to where these extra branches occur in figure 2 (c) and (d), in a manner suggested in [16, 19], may allow their experimental observation. The minimum that the system will be trapped in depends on the rate of cooling. With reasonably quick cooling rates and large surrounding energy barriers, the system may be trapped in a local minimum above the absolute minimum with the desired level of magnetisation.

In figure 3 (a) the critical lines of stability in the $H_x - H_y$ plane are shown for the value $K = 0.7$. These critical lines mark the points in the energy landscape where there is an energy barrier separating the local minima. Each of the barriers is associated with a saddle point. Large broad minima are more likely to be occupied than shallow minima. This is due to the larger basin of attraction of the deeper wells. The critical points in the magnetisation curves of figure 2 lie at the points where there is a discontinuity or a continuous phase transition. These critical points correspond to changes in the energy landscape which occur as the magnetisation vectors evolve with a changing magnetic field strength. Figure 3 (b) to (e) shows the stable minima magnetisation curves of figure 2 with the critical points clearly shown. How these vectors switch at a discontinuity or continuous phase change is shown in figure 3 (f) to (i). Considering the four interacting particles to have a domain structure, the lowest energy configuration at $H = 0$ occurs when the magnetisation vectors are aligned anti-parallel to one another such that they can be described as antiferromagnetic. This means that the self magnetic fields of the system cancel and so the system has no net magnetic field. This only occurs for even numbers of particles in the system. When the magnetic particles are in a strong enough external magnetic field, the domains re-orientate in parallel with the field. This happens as a continuous phase transition in figure 3 (b)-(e) at critical points $C_m^\beta = C_6^0, C_5^{18}, C_3^{54},$ and $C_5^{72}$, respectively. Here m is the index referring to the $m^{th}$ phase transition as the magnetic field strength, H, increases from zero.

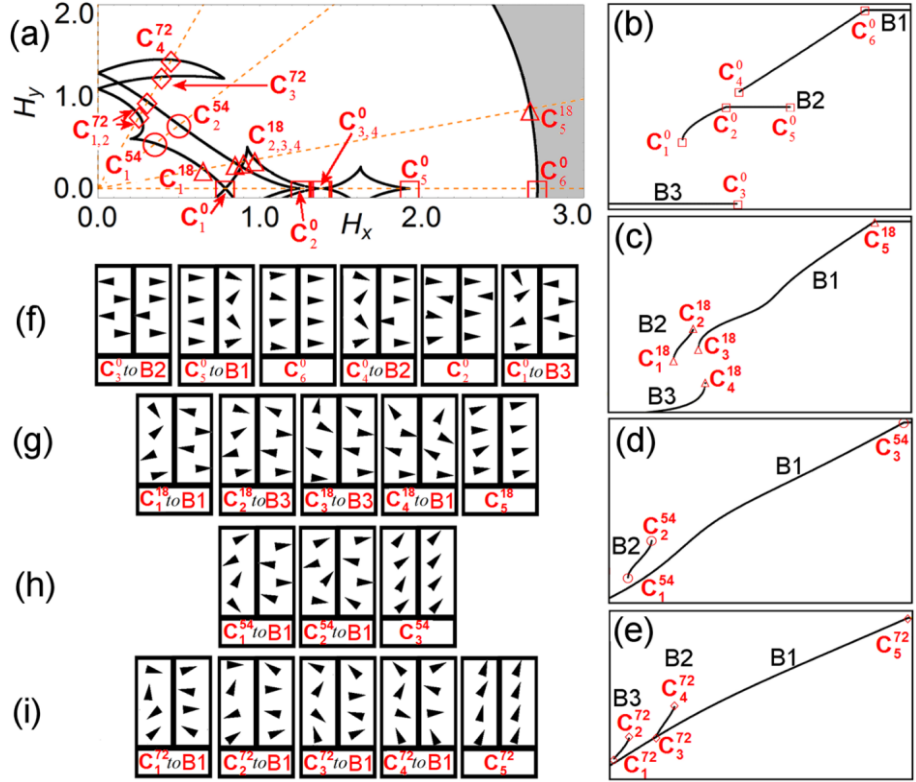

*Figure 3. For* $K = 0.7$ *the critical points that mark the evolution of the energy minima of the system are shown. (a) All areas of the* $H_x - H_y$ *phase diagram are stable minima. The critical points on the critical lines of stability are shown at* $\beta = 0°, 18°, 54°, 72°$. *The magnetisation branches of Figure1 are reproduced for* $H > 0$ *in (b) to (e) showing the critical points that mark phase changes in the system. In (b) the phase changes occur at* $H = 0.788, 1.252, 1.375, 1.386, 1.928$ *and* $2.714$ *in the order of* $C_{1 \to 6}^0$. *In (c)* $H = 0.686, 0.893, 0.948, 1.02,$ *and* $2.807$ *in the order of* $C_{1 \to 5}^{18}$. *For* $C_{1 \to 3}^{54}$ *in (d) the critical field values occur at* $H = 0.641, 0.855$ *and* $3.512$. *The critical points* $C_{1 \to 5}^{72}$ *in (e) correspond to* $H = 0.831, 0.994, 1.279, 1.473,$ *and* $3.934$. *In the phase diagram in (a) these critical field points are found through the relations* $H_x = H\cos\beta$ *and* $H_y = H\sin\beta$. *Boxes (f) to (i) show how the magnetisation vectors in the four particles orientate themselves at these critical points. The notation* $C_m^\beta$ *is for the* $m^{th}$ *critical point at applied field angle* $\beta$. *In the boxes of (f) to (i) showing the magnetic moment orientations, those orientations that are underlined by a* $C_m^\beta$ *to B statement are discontinuous transitions whereby a relaxation or jump from the critical point to some branch of the magnetisation curves is made. Those that are underlined simply by* $C_m^\beta$ *statements show the phase states just before and after a continuous phase change.*

In figure 3 (b) – (e) the interdependence of the anisotropy, magnetic field strength and the angle of the applied magnetic field is shown. The system can be very susceptible to small changes of these parameters. The rearrangements of the domain structure at the critical points can involve localized transitions or involve the whole system. For example, in Figure3 (f) the transition from $C_3^0$ to $B2$ is a discontinuous jump that happens with three of the magnetisation vectors remaining in the same orientation and one going through a spin flip from $\varphi_1 = 180°$ to $0°$. However, many of the discontinuities involve more than one particle changing its orientation and frequently involve all four. Zero remanence is seen in all of the magnetisation curves that are functions of the applied magnetic field. This is a typical feature of systems of magnetic particles of even numbers. This makes these systems ideal candidates for bio-applications as the disappearance of the magnetisation reduces the long range magnetostatic forces responsible for agglomeration of the particles [15].

## 3. Apoptosis

Recently it has been proposed to use spinning magnetic disks in the vortex state to "cut" the cellular membrane and initiate apoptosis [15]. It has also been shown recently that chains of magnetic particles coated in dextran can evade the defensive mechanisms of the body for long periods of time [3]. This is due to elongated structures making internalization by phagocytosis more difficult to occur. Here a new tool for targeting and combating cancer-cells is outlined. An array of stable single domain magnetic particles, separated by non-magnetic spacer layers could potentially make it extremely difficult for unwanted cellular uptake to occur. As previously mentioned, the vector at which an elongated particle approaches a macrophage determines how long it takes for internalization to occur [4]. This is analogous to a snake trying to consume its prey. It does so along the length of the body. Thus, a multilayer of elongated magnetic particles presents a difficult object, along all axes, for internalization to happen. It is designed to have high aspect ratios in each direction.

Figure 4 illustrates the nanoblade concept. Suitably structured SAF multilayers could circulate in the body and target cancer cells very effectively. Recently Zhang et al successfully targeted brain tumours with a magnetic core nanoparticle [18]. The nanoparticle crossed the blood brain barrier with a biocompatible coating of copolymer to which a tumour-targeting agent was attached. Therefore, the nanoparticle is capable of selectively targeting tumours. Once a nanoblade has found the tumour, a low magnetic field can be used to initiate its rotation. The mechanical force against the tumour then exposes the cancer cell to a localized simulation that is strong enough to begin apoptosis. The required force to do this has been found to be on the order of a few pico-newtons [15].

Apoptosis, or programmed cell death (see figure 4), is a regulated process in the body that eliminates dysfunctional cells [20]. It allows the cells to degrade in a clean, controlled process that restricts the content of the cell to remain inside the cell membrane as it is destroyed. Macrophages, by means of phagocytosis, remove the apoptotic cell before the cell can release any of its content. It has been suggested by Kim et al [15] that magneto-mechanical cellular membrane disturbance can initiate calcium signaling, which begins the apoptosis reaction. An influx of calcium into the mitochondrion or the binding of the Bid protein to the mitochondrial membrane heralds the death stimuli of the cell. A permeability transition is induced into the membrane of a neighboring mitochondrion, resulting in the release of cytochrome C. This is the protein that initiates cell death and is present in almost all forms of life. In healthy cells, the mitochondria are coated with the protein Bcl-2, which inhibits apoptosis. Upon damage to the cell Bax and Bid proteins are released from inside the mitochondria and move to its surface. These proteins inhibit the effect of Bcl-2 and together with another apoptotic protein, Bak, create pores in the outer mitochondrial membrane. This culminates in the release of cytochrome C. The creation of these pores is the pivotal element in programmed cell death and marks the beginning of the end of the cell. With the release of the cytochrome C a complex is formed with the high-energy molecule adenosine triphosphate (ATP) and the enzyme Apaf-1. The creation of this complex in the cycoplasm stimulates the initiator protein caspace-9. A multi-protein complex apoptosome is now formed through the interaction of caspace-9, cytochrome C, ATP and Apaf-1. Caspace-3 is now activated by the apoptosome to begin degradation. Apoptosis inducing factor (AIF) is released to fragment DNA as well as smac/Diablo proteins to prevent the inhibition of apoptosis. The programmed cell death, stirred into action by the nanoblade, now works as it should have in the first place to prevent the cancer cells becoming "immortal". When apoptosis fails to function in the body, as a natural method of cell clearance, cancers can emerge [20]. Biological skin markers of cancer and the dependency on glutamine was also discussed by the author, and it was suggested that formations associated with glutamine could also assist in bio-functionalising micro and nanoparticles for high efficacy [21]. The dynamics and nanomechanics of nanomagnets in vivo were discussed in reference [22].

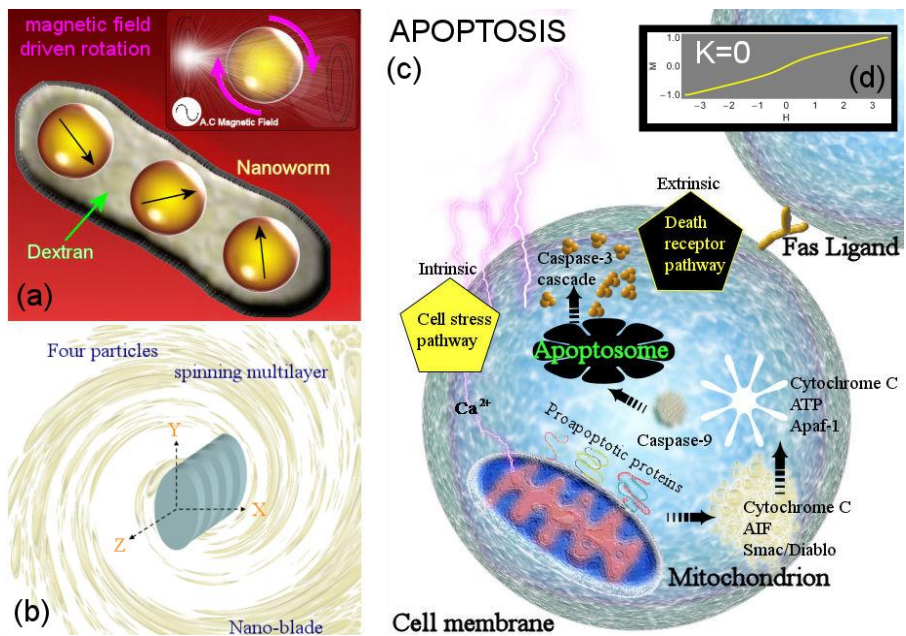

*Figure 4. (a) Nano-worms [3] consisting of magnetic particle cores are assembled in an elongated chain. Magnetic particles become aligned along strands of dextran. Phagocytes readily take up spherical geometries, whereas elongated particles have a prolonged blood half-life. (b) It is proposed that a chain or multilayer grouping of an even number of magnetic particles could act as a nano-blade to initiate apoptosis. (c) The initiation of programmed cell death, apoptosis, by the spinning motion of a nano-blade. (d) The net magnetisation of a nanoblade consisting of multilayers of circular discs, with $K = 0$, shows that there is no remanence or coercivity. This is the case for all geometries consisting of even numbers of elliptical disks.*

**4. The force generated by a magnetic field on a nanoblade**

The nanoblade consists of a series of N magnetic disks. The stable single domain multilayers are separated by some interlayer composed of non-magnetic material. Gold could be used for this purpose [1], or ruthenium [2]. The multilayer can also be coated in gold to ensure biocompatibility. A gold outer layer allows the surface of the nanoblade to be further coated in cancer-targeting ligands [1]. In figure 5 a comparison of the discussed magnetic response against the experimental results of Hu et al [8] is given for systems of two and four high-moment disks. In order to simulate the magnetisation as a function of the applied magnetic field, experimental deviations in the synthesis of the SAF geometry has to be considered. In the experiments the SAF are almost circular flat disks. As such, when the magnetic disks are produced the uniaxial anisotropy in each disk may be offset from each other. This explains the "rounded" shape of the M versus H plots. To account for this the magnetic field applied to the SAF is taken to propagate at angle $\beta_1$ to the first particles easy axis. The second and consequent magnetic disks experience an offset of $\beta_i = \beta_1 + \xi_i$. Owing to the shape variations each magnetic disk in the SAF may have a slightly differing anisotropy constant, K. In figure 5 (a) the two magnetic disks have anisotropy $K = 0.02$ and the applied magnetic field angles are at $\beta_1 = 75°$ and $\beta_2 = 85°$. This offset in the easy-axes of the two disks has the effect of changing the appearance of the magnetisation curve from a mostly "square" shape when $\beta_1 = \beta_2$ to the rounder shape of the experiment. Figure 5 (b) is the comparison of the model for a four magnetic disk SAF against the experimental data[8]. The values of anisotropy are $K_{1-4} = 0.01, 0.03, 0.05, 0.07$ and the applied field angles are taken to be $\beta_1 = \beta_3 = 45°$ and $\beta_2 = \beta_4 = 35°$. This results in a good comparison of the model to the experimental data.

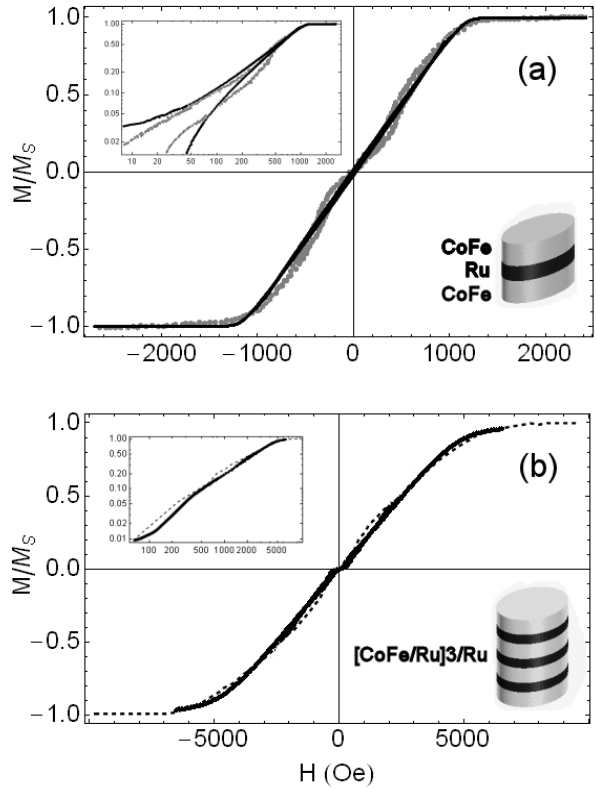

*Figure 5. A comparison of the experimental results in Hu et al [8] figure 2 to the theoretical model. The dotted grey lines are the experimental values and the black solid lines are from the model. The easy axes of each particle in the stack are slightly offset against one another and fabrication deviations are considered. The top and bottom plots are for two and four magnetic disk SAF systems. The insets in each left hand corner are $(\text{Log}(M/M_s), \text{Log} H)$ plots.*

Knowing that the model can produce accurate results, an evaluation of the force on an illustrative four magnetic disk nanoblade is now made. Figure 6 shows the net force acting on two SAF stacks of elliptical disk structures [CoFe(5)/Ru(0.8)]3/CoFe(5) and a [CoFe(5)/Au(10)]3/CoFe(5). Each elliptical magnetic disk, as a single domain ferromagnetic particle, is considered in analogy to a bar magnet. At $H = 0$, the magnetisation vectors are aligned in anti-parallel, $\varphi_1 = \varphi_3 = 0$ and $\varphi_2 = \varphi_4 = 0$. In this case there is no net torque, on the nanoblade. The net torque is also zero when the magnetic field is applied at $\beta = 90°$ as a consequence of there being an even number of magnetic particles in the system.

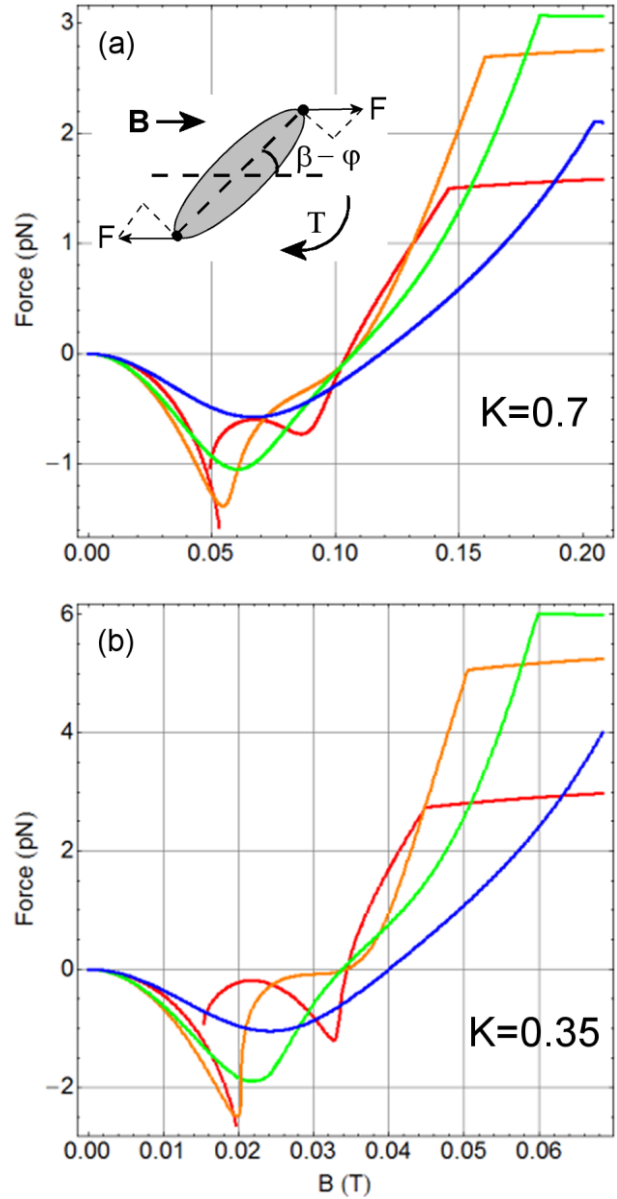

Figure 6. (a) The net force acting on a [CoFe(5)/Ru(0.8)]3/CoFe(5) SAF stack of elliptical disks at various angles of applied magnetic flux density $\mathbf{B}$. The red lines are $\mathbf{B}$ at $18°$, the orange at $36°$, the green at $54°$ and the blue at $72°$ to the easy axes of the magnetic particles in the stack. The inset shows a top down view of the magnetic torque acting on the magnetic particles. The magnetic particles constituting the SAF structure have semi-axis lengths of $a = 25$nm and $b = 10$nm. (b) The same as in (a) but with SAF structure [CoFe(5)/Au(10)]3/CoFe(5), $a = 50$nm and $b = 46$nm.

The torque is found through $\mathbf{T} = \mathbf{m} \times \mathbf{B}$, where $\mathbf{m}$ is the magnetic moment. This results in an equation for the net torque on the nanoblade,

$$T = B \sum_i^N M_{Si} v_i \sin(\beta_i - \varphi_i) \qquad (4)$$

The net force acting on the nanoblade is found as $F = T/a$. The force is a function of the applied magnetic field orientation and the magnetisation vector angle. It can be seen that for bio-functionality the prerequisite that the system be smaller that 100nm and preferentially smaller than 50nm is attainable. This is because the force generated meets the pN range required to initiate apoptosis. The net force that the nanoblade can produce can be tailored by the geometry of the magnetic disk. As we have seen in figures 1 - 4, the energy landscape associated with the nanoblade is complicated and varying as a function of its geometry. The ellipticity of the magnetic disk cross section, the separation distance between disks, and the number of magnetic disks dictate the form of the anisotropy and interaction energy. Introducing more magnetic disks into the SAF will generate a greater torque in response to the applied magnetic field. Also, the Barkhausen jumps seen in the magnetisation as the magnetic field is swept can be used to give a small extra nudge to the cancer. At these discontinuities there is a very fast change in the phase of the system, which is marked by the magnetisation vectors rotating very suddenly into another orientation.

## 5. Conclusion

To summarize, the magnetic response of a nanoblade to an applied magnetic field has been shown to exhibit a complicated series of phase transitions in the energy landscape. The nanoblade operates in the regions of this energy landscape that correspond with stable energy minima. The nanoblade has been illustrated for a system of four magnetic disks separated by non-magnetic interlayers. The benefit of scaling it up to include N magnetic disks would be an increased torque capability. However, as N increases so does the system complexity. There can be many hysteresis paths in a large system of magnetic particles and prediction of the hysteresis path that the system follows becomes less predictable (see [16]). Therefore, we argue that nanoblades that consist of four to six magnetic disks are most optimal for surgical operations.

Also, the system should be designed to start apoptosis with sufficient but not excessive force. Too much force may initiate necrosis which is usually brought about by an external trauma to the cellular membrane. Gently starting apoptosis allows the body to use its natural cellular clearance methods. The nanoblade holds to the principle of trying to reduce invasive surgical techniques and work with the body to cure itself. These high moment devices can be actuated in the vicinity of the cancerous tissue with very low magnetic fields. Thus, the nanoblade can be operated without detriment to the normal physiological processes of the body due to its high sensitivity to the applied field. We state that it offers a viable alternative to chemotherapy or radiotherapy without the long term side-effects. We expect that the proposed nanoblade can be a very essential tool for cancer treatment in the near future.

**References**
[1] Punnakitikashem. P; et al. Design and fabrication of non-superparamagnetic high moment magnetic nanoparticles for bioapplications. J Nanopart Res, **2010**, 12, 4, 1101-1106.
[2] Hu. W., Synthetic antiferromagnetic nanoparticles with tunable susceptibilities, Journal of Applied Physics, **2009**, 105, 07B508.
[3] Park. J-H.; et al. Magnetic iron oxide nanoworms for tumour targeting and imaging. Adv. Materials, **2008**, 20, 1630-1635.

[4] Mitragoti. S.; Lahan. J. Physical approaches to biomedical design. Nature Materials, **2008**, 8, 15-23.
[5] Champion. J. A.; Mitragoti. S. Role of target geometry in phagocytosis. PNAS, **2006**, 103, 13, 4930-4934.
[6] Hoffmann. A. Magnetic viruses for biological and medical applications. Magnetic business and technology magazine, Spring **2005.**
[7] Doshi. N.; Mitragoti. S. Designer biomaterials for nanomedicine. Adv Funct. Mater, **2009**, 19, 3843-3854.
[8] Hu. W.; et al. High moment antiferromagnetic nanoparticles with tunable magnetic properties. Adv. Materials, **2008**, 20, 1479-1483.
[9] Wang. S-Y.; Fujiwara.H.; Sun. M. Bias field effects on the toggle mode magnetoresistive random access memory. Journal of Applied Physics, **2006**, 08N903. 14
[10] Forrester. D. M.; Kürten. K. E,; Kusmartsev. F. V. Two particle element for magnetic memory. Phys Rev B, **2007**, 134404.
[11] Plamada. A-V.; Cimpoesu. D.; Stancu. A. Activation energy and switching behaviour of two identical magnetic particles. Applied Phys Letters, **2010**, 122505
[12] Forrester. D. M.; Kovacs. E.; Kürten. K. E,; Kusmartsev. F. V. Astroid curves for a synthetic antiferromagnetic stack in an applied magnetic field. Int. J. of Mod. Phys. B, **2009**, 232, N. 20-21, 4021-4040
[13] Konovalenko. A.; et al. Spin dynamics of two-coupled nanomagnets in spin-flop tunnel junctions. Phys Rev B, **2009**, 80, 144425
[14] Forrester. D. M.; Kürten. K. E.; Kusmartsev F. Fundamental design paradigms for systems of three interacting magnetic nanodiscs. Appl. Phys. Lett. **2011**, 98, 163113
[15] Kim. D-H.; et al. Biofunctionalized magnetic-vortex microdiscs for targeted cancer cell destruction. Nature Materials, **2010**, 9, 165-171.
[16] Forrester. D. M.; Kürten. K. E.; Kusmartsev. F. V. Magnetic cellular automata and the formation of glassy and magnetic structures from a chain of magnetic particles. Phys Rev B, **2007**, 014416
[17] Beleggia. M.; et al. Demagnetisation factors for elliptic cylinders. J. Phys.D: Appl. Phys. **2005**, 38, 3333-3342.
[18] Zhang. M.; et al. Specific targeting of brain tumours with an optical/magnetic resonance imaging nanoprobe across the blood brain barrier. Cancer Research, **2009**, 69, 15, 6200-6207
[19] Forrester. D. M.; Kürten. K. E.; Kusmartsev. F. V. Fractal metamaterials composed of electrically isolated π-rings. Sci. Lett. J. **2015**, 4: 133.
[20] Heemels. M-T., editor. Apoptosis. Nature, **2000**, 407, 770-810.
[21] Forrester. D. M. Self-assembled multi-ring formations of glutamine and a possible link to erythema gyratum repens. Medical Hypotheses **2015**. 85,1, 10-16.
[22] Forrester. M.; Kusmartsev. F. The nano-mechanics and magnetic properties of high moment synthetic antiferromagnetic particles. Physica Status Solidi a: Applications and Materials Science 2014. 211, 4, 884-889.